\documentclass[12pt]{article}

\usepackage{scicite}
\usepackage{graphicx}
\usepackage{times}
\usepackage{xcolor}
\usepackage{amsfonts}
\usepackage{esvect}
\usepackage{harpoon}
\usepackage{amsmath,accents}
\newcommand{\myvect}[1]{\accentset{\rightharpoonup}{#1}}

\newenvironment{sciabstract}{%
\begin{quote} \bf}
{\end{quote}}

\title{Metasurface-Based Realization\\ of Photonic Time Crystals}

\author
{Xuchen Wang,$^{1,2\ast}$~Mohammad Sajjad Mirmoosa,$^{1}$ ~Viktar S. Asadchy,$^{1,3}$\\ ~Carsten Rockstuhl,$^{2,4}$
~Shanhui Fan,$^{3}$ ~Sergei A. Tretyakov$^{1\ast}$
\\
\normalsize{$^{1}$Department of Electronics and Nanoengineering, Aalto University, Espoo, Finland}\\
\normalsize{$^{2}$Institute of Nanotechnology, Karlsruhe Institute of Technology, Karlsruhe, Germany}\\
\normalsize{$^{3}$Ginzton Laboratory and Department of Electrical Engineering, Stanford University, USA}\\
\normalsize{$^{4}$Institute of Theoretical Solid State Physics, Karlsruhe Institute of Technology, Karlsruhe, Germany}\\
\\
\normalsize{$^\ast$Corresponding author. E-mail:  xuchen.wang@kit.edu; sergei.tretyakov@aalto.fi;}
}

\date{}

\begin{document} 


\baselineskip24pt


\maketitle


\begin{sciabstract}
Photonic time crystals are artificial materials whose electromagnetic properties are uniform in space but periodically vary in time. The synthesis of such materials and experimental observation of their physics remain very challenging due to the stringent requirement for uniform modulation of material properties in volumetric samples.
In this work, we extend the concept of photonic time crystals to two-dimensional artificial structures -- metasurfaces.
We demonstrate that time-varying metasurfaces not only preserve key physical properties of volumetric photonic time crystals despite their simpler topology but also host common momentum bandgaps shared by both surface and free-space electromagnetic waves.  
Based on a microwave metasurface design, we experimentally confirmed the exponential wave amplification inside a momentum bandgap as well as the possibility to probe bandgap physics by external (free-space) excitations. The proposed metasurface 
serves as a straightforward material platform for realizing emerging photonic space-time crystals and as a realistic system   for the amplification of surface-wave signals in future   wireless communications.
\end{sciabstract}
\noindent
Teaser:  Photonic time crystals become thinner


\section*{Introduction}
Time, as an additional degree of freedom, has substantially extended the potential of artificial electromagnetic materials \cite{ galiffi2022photonics, shaltout2019spatiotemporal, engheta2021metamaterials}. In the past few years, numerous new physical effects have been discovered based on time modulation of material properties, such as magnetless nonreciprocity \cite{yu2009complete, estep2014magnetic}, effective magnetic field for photons~\cite{fang2012realizing,tzuang2014non}, synthetic dimensions \cite{lustig2018topological, dutt2020single, leefmans2022topological, wang2021topological}, electromagnetic devices beyond physical bounds \cite{10041963}, among many others \cite{galiffi2022photonics}. 
One of the major developments in this direction is the concept of photonic time crystals (PhTCs) \cite{martinez2016temporal,lyubarov2022amplified}.
PhTCs are artificial materials whose electromagnetic properties (such as permittivity or permeability) are periodically and rapidly modulated in time while remaining uniform in space. They are temporal counterparts of spatial photonic crystals~\cite{joannopoulos2008molding}. In analogy to frequency bandgaps in spatial photonic crystals, the temporal modulation of PhTCs results in momentum bandgaps. Due to the non-Hermitian nature of PhTCs, inside the momentum bandgap, the electromagnetic wave grows exponentially in time~\cite{lustig2018topological}. This is in sharp contrast to wave attenuation in time occurring inside a frequency bandgap~\cite{joannopoulos2008molding,sievenpiper1999high}. 
Recently, numerous interesting wave phenomena have been theoretically predicted in PhTCs, such as topologically nontrivial phases~\cite{lustig2018topological}, a temporal analog of Anderson localization~\cite{carminati2021universal}, amplified radiation from free electrons and atoms~\cite{dikopoltsev2022light,lyubarov2022amplified}, etc. 

The realization of PhTCs typically implies temporal modulations  of material properties in the \textit{bulk} (inside a three-dimensional volume). At the lower frequencies, temporal modulations of material properties are usually achieved by variable electronic components such as varactors~\cite{hadad2016breaking}. Implementing three-dimensional PhTCs with varactors would require  a  structure with overly complex pumping network (the latter would additionally lead to parasitic interferences with electromagnetic waves propagating in the bulk). Therefore, previous studies at microwaves relied on \textit{mimicking} spatially infinite structure with circuits~\cite{reyes2015observation} and closed waveguides~\cite{park2022revealing}. However, despite their fundamental importance, these toy models of PhTCs had no potential for practical applications due to their \textit{closed-system} nature (no interaction with free-space waves).
%
%
In the optical frequencies, temporal modulations are usually obtained by ultrafast carrier accumulation and depletion that is based on strong light pumping. 
Due to the spatial non-uniformity of the pump light, achieving clean PhTC regime becomes very complicated. There have been several important developments toward the implementation of optical PhTCs  based on epsilon-near-zero materials~\cite{kinsey2015epsilon,vezzoli2018optical, zhou2020broadband}, but till now no successful realization has been reported.

In this work, we  introduce the concept of metasurface-based PhTCs. Such   crystals have   an extent   in one temporal and only two spatial    dimensions (along which the metasurface is uniform). This  reduction of dimensionality   allows us to overcome the aforementioned implementation problems   and propose a very practical design of PhTCs.  
Similar dimensionality reduction     played  an enormous   role in   electromagnetics and photonics in the past,  leading to notably more feasible systems~\cite{kildishev2013planar} (transition from three-dimensional metamaterials to two-dimensional metasurfaces accompanied by game-changing reduction in cost, attenuation loss, and complexity)  and new systems
with drastically different applications~\cite{sievenpiper1999high} (extension of photonic crystals to  two-dimensional mushroom-type   surfaces with numerous applications for the
antenna industry).
Our metasurface-based PhTC, in addition to straightforward implementation, enables   richer physics compared to that of bulk PhTCs.
In particular, the proposed metasurface gives rise to momentum bandgaps shared by both surface and free-space propagating waves. Due to the Floquet coupling of the frequency harmonics, we were able to probe the momentum bandgap in a PhTC by its direct excitation from free space. In such a setting,    the required modulation frequency is substantially reduced compared to the signal frequency. Moreover, we  experimentally confirmed exponential wave amplification inside the photonic momentum bandgap. Due to simple two-dimensional geometry, the proposed metasurface platform can be used to design emerging space-time crystals~\cite{caloz2019spacetime,  sharabi2022spatiotemporal} and in future communication systems operating with surface waves~\cite{shojaeifard2022mimo}. While as proof of concept, we designed metasurface PhTC for the microwave band, the idea can be further extended to metasurface configurations operating at higher frequencies (based on graphene or 2D materials).
Finally, we highlight a qualitative difference between PhTCs and degenerate parametric amplifiers.

\section*{Results}
\subsection*{Eigenmode analysis} 

\begin{figure*}
\includegraphics[width=0.99\linewidth]{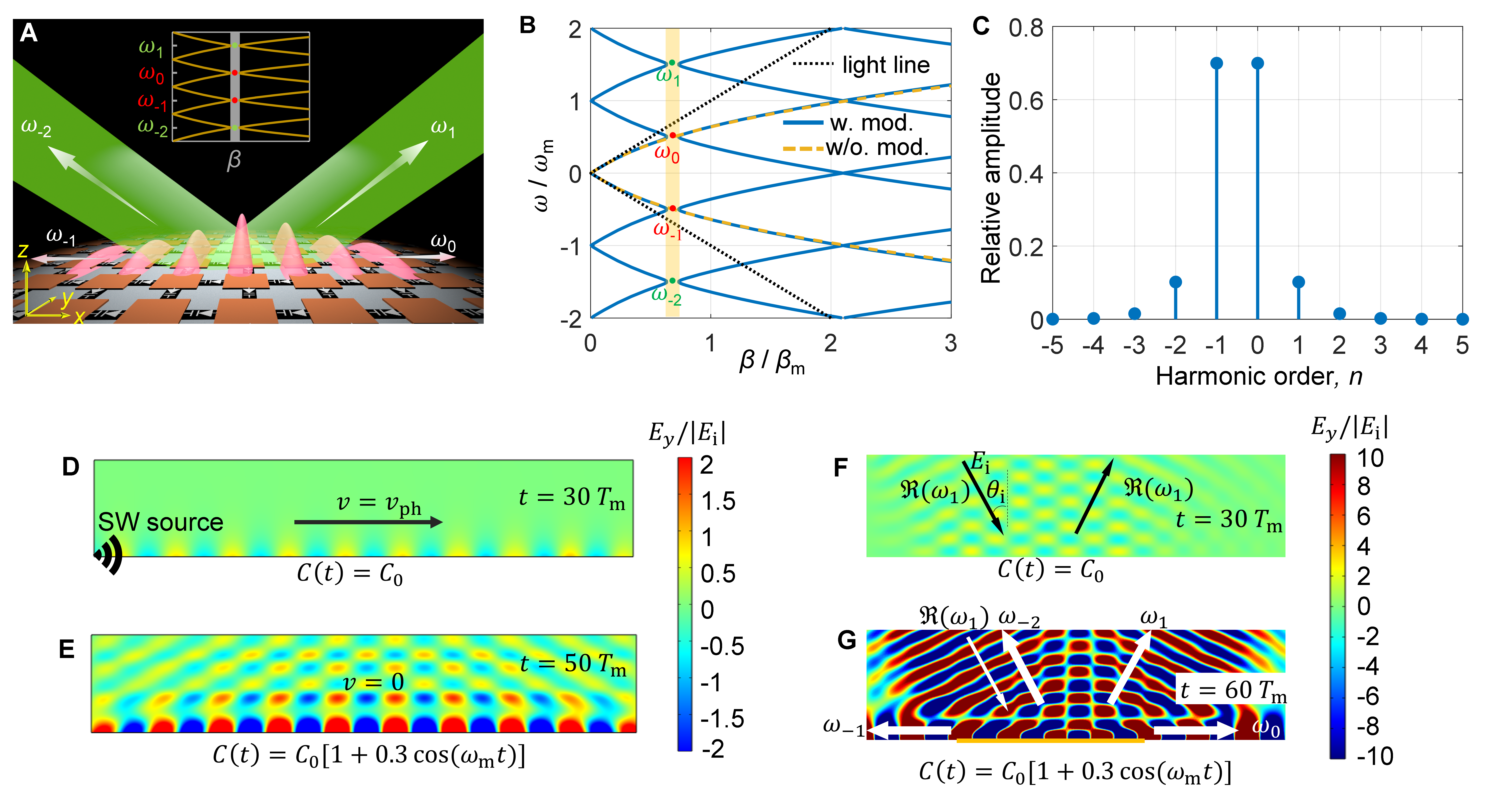}
\caption{\textbf{Theoretical and numerical results.} (A) Conceptual picture of the metasurface PhTCs. 
The momentum bandgap (grey gap in the inset) opens when the surface properties are  uniformly modulated in time. In the bandgap, multiple harmonics are symmetrically excited with different frequencies but the same wavevector.
(B) Band structure (blue solid line) and bandgap (orange area) of the time-varying capacitive boundary. The dispersion curve of the time-invariant boundary (brown dashed line) is derived in section S4. 
The horizontal axis depicts the wavevector normalized to $\beta_{\rm m}=\omega_{\rm m}\sqrt{\epsilon_0\mu_0}$. The band diagram is periodic along the frequency axis. (C) Spectral composition of electric-field harmonics in the eigenmodes inside the momentum bandgap with $\beta_{\rm g}=0.69\beta_{\rm m}$. 
(D) Electric field distributions above the boundary at $t=30T_{\rm m}$ and (E) $t=50T_{\rm m}$.
The fields are normalized to the surface wave excitation amplitude $|E_{\rm i}|$. (F) Electric field distribution above the boundary under plane wave incidence at $t=30T_{\rm m}$ and (G) $t=60T_{\rm m}$. }
\label{fig: bbigfigure}
\end{figure*}

Figure~\ref{fig: bbigfigure}A shows a spatially homogeneous metasurface at the $z=0$ plane. The metasurface is modeled as a capacitive impenetrable surface impedance boundary~(section S1). 
The effective capacitance of the boundary is periodically modulated in time as $C(t)=C(t+T_{\rm m})$, where $T_{\rm m}$ is the modulation period. An arbitrary periodic modulation function can be expressed as a sum of Fourier terms, $C(t)=\sum_p c_p e^{jp\omega_{\rm m}t}$, where $p\in \mathbb{Z}$ is the Fourier index, $c_p$ are the Fourier coefficients, and $\omega_{\rm m}=2\pi/T_{\rm m}$ is the modulation frequency. Here, we assume that fields vary in time according to ${\rm e}^{j \omega t}$ convention. Next, we use the eigenmode analysis (similar to that in~\cite{zurita2009reflection}) to extract the band structure of this time-varying boundary.
Assume that TE-polarized eigenwaves propagate along the $x$-direction. 
The temporal modulation induces an infinite number of frequency harmonics $\omega_n=\omega_0+n\omega_{\rm m}$, where $\omega_0$ is the frequency of the fundamental harmonic and $n$ is the Floquet order. For each value of wavenumber $\beta$, the tangential electric (along $y$-direction) and magnetic fields (along $x$-direction) of the eigenmode are expressed as a sum of these harmonics:
\begin{subequations}
\begin{align}
\myvect{E}=\sum_n E_ne^{-\alpha_nz}e^{-j(\beta x-\omega_n t)}\hat{y}, \\ \myvect{H}=\sum_n H_ne^{-\alpha_nz}e^{-j(\beta x-\omega_n t)}\hat{x},
\end{align}\label{eq1}
\end{subequations}
where $\alpha_n$ denotes the attenuation constants along the $z$-direction regarding each harmonic, $E_n$ and $H_n$ are spectral amplitudes related by wave admittance, i.e., $H_n=({\alpha_n}/{j\omega_n\mu_0})E_n$. Since every harmonic must satisfy the Helmholtz equation in free-space, we obtain the condition $\beta^2=\alpha_n^2+\omega_n^2\epsilon_0\mu_0$. Additionally, by substituting Eq.~(\ref{eq1}) into the time-domain boundary condition $\int \hat{z}\times \myvect{H} dt=C(t)\myvect{E}$ of a capacitive surface (section S1), we also obtain (section S2)
\begin{equation}
    \sum_p j\omega_n c_p E_{n-p}=H_n. \label{eq: harmonic relation 2}
\end{equation}

The spectral content of the harmonics is truncated with Floquet orders from $n=-N$ to $n=+N$. Because all the considered harmonics must satisfy Eq.~(\ref{eq: harmonic relation 2}), we have in total $2N+1$ equations that are written in a matrix form $\mathbf{Y}\cdot \mathbf{E}=\mathbf{H}$. Here, $\mathbf{Y}$ is a square matrix related to the Fourier coefficients~$c_p$ and frequencies~$\omega_n$, $\mathbf{E}$ and $\mathbf{H}$ are column vectors containing harmonics $E_n$ and $H_n$, respectively. By combining this matrix equation with $\mathbf{M}\cdot\mathbf{E}=\mathbf{H}$ that relates the electric and magnetic field column-vectors through the admittance matrix $\mathbf{M}$, we finally deduce the following equation: 
\begin{equation}
\big[\mathbf{Y}-\mathbf{M}\big]\cdot\mathbf{E}=0. 
\label{eigenvalue}
\end{equation}
The definitions of the matrices are shown in supplementary section S3. Thus, according to the above expression, the dispersion relation for waves propagating along time-varying capacitive boundaries is given by $\det\big[\mathbf{Y}-\mathbf{M}\big]=0$, in which ``$\det$'' denotes the determinant.

As a specific example, we consider a boundary whose effective surface capacitance is mo\-du\-lated according to $C(t)=C_0[1+0.3\cos(\omega_{\rm m}t)]$. 
The median surface capacitance is expressed through the surface admittance as $C_0=Y_{\rm s}/j \omega_{\rm 0}$, where $Y_{\rm s}=j2.5 \times 10^{-3}$~S is assumed. The band diagram for this case is depicted with a blue line in Fig.~\ref{fig: bbigfigure}(B). Analogously to the space-modulated boundaries where frequency bandgaps occur~\cite{sievenpiper1999high, joannopoulos2008molding}, for the time-modulated boundary, the corresponding bandgap is generated in the $k$-space, as the figure confirms. Inside the momentum bandgap, frequencies $\omega$ are complex-valued.

\subsection*{Wave propagation inside a momentum bandgap}
It is also possible to determine eigenmodes inside the bandgap by solving equation (\ref{eigenvalue}) with respect to the frequency. 
As an example, we choose the wavenumber in the center of the momentum bandgap $\beta_{\rm g}=0.69\beta_{\rm m}$. 
The corresponding eigenfrequencies (see the colored points in Fig.~\ref{fig: bbigfigure}(B)) are complex values given by $\omega_n= (0.5+n\pm j0.023)\omega_{\rm m}$, where $+j$ and $-j$ indicate that the harmonic exponentially decays or grows in time, respectively. All these frequency harmonics exist simultaneously with specific amplitudes and phases that satisfy equation (\ref{eigenvalue}). The spectral content $E_n$ of the growing and decaying eigenmodes for $\beta_{\rm g}$ is shown in Fig.~\ref{fig: bbigfigure}C. Since the decaying harmonics eventually disappear as time passes, we only consider amplified harmonics hereafter.

It is observed that the harmonic amplitudes are symmetrically distributed in the spectrum. The  $n=0$ and $n=-1$ harmonics are dominant and propagate along the capacitive boundary with equal amplitudes, but the real parts of their frequencies have opposite signs. These two harmonics have opposite phase velocities  $v_{\rm ph}=\Re(\omega)/\beta$ and together form a standing surface wave with a  temporally growing amplitude. 
Importantly, the higher-order harmonics ($n>0$ and $n<-1$) for the considered scenario are outside the light cone and correspond to free-space propagating waves. This means that the temporal modulation not only excites surface harmonics but additionally results in  the coupling of surface harmonics to free-space propagating harmonics, and all these harmonics have the same amplification rate.

Next, we numerically verify the wave evolution in the momentum bandgap. First, we consider the scenario where the momentum bandgap is probed by surface-wave excitations. From the time $t=0$ until $t=30 T_{\rm m}$ ($T_{\rm m}=2\pi/\omega_{\rm m}$), a surface wave is launched onto the stationary (or static) capacitive boundary $C(t)=C_0$ with $\omega=0.5\omega_{\rm m}$ and $\beta=0.69\beta_{\rm m}$ from the left simulation port located above the boundary (see Fig.~\ref{fig: bbigfigure}D). After $t>30T_{\rm m}$, the temporal modulation of the capacitive boundary is turned on. Since the modulation occurs in time and the structure remains uniform in space, the momentum of the waves remains unchanged. However, the modulation generates backward and forward surface harmonics with equal amplitudes (see Fig.~\ref{fig: bbigfigure}C), resulting in a standing wave along the horizontal direction with zero group velocity, as shown in Fig.~\ref{fig: bbigfigure}E. 
Together with the surface harmonics, free-space harmonics are symmetrically generated and exponentially amplified in time (according to Fig.~\ref{fig: bbigfigure}C). The complete field evolution animation is available in movie S1. 

Furthermore, since, according to Fig.~\ref{fig: bbigfigure}B, the momentum gap splits bands corresponding to both surface waves and free-space waves, it is feasible to access the momentum gap by obliquely illuminating the boundary with an \textit{external} plane wave at the tangential wavevector $\beta_{\rm g}$ and frequency $\Re(\omega_1)$. In a second set of simulations (movie S2), from the time $t=0$ until $t=30 T_{\rm m}$, we launch onto the time-invariant boundary such free-space plane wave at the incident angle $\theta_{\rm i}=\arcsin[{c\beta_{\rm g}}/{\Re(\omega_1)}]$ (see Fig.~\ref{fig: bbigfigure}F). The reactive boundary fully reflects the incident wave, forming a standing wave pattern in the normal direction. The temporal modulations of the boundary are turned on after $t=30T_{\rm m}$ in a finite region (marked by orange color in Fig.~1G). The temporal modulation excites surface harmonics ($\beta_{\rm g}, \omega_{0}$) and ($\beta_{\rm g}, \omega_{-1}$), as well as plane-wave harmonics which include ($\beta_{\rm g}, \omega_{1}$) and ($\beta_{\rm g}, \omega_{-2}$). All these harmonics experience exponential growth at the same rates. Therefore, the boundary acts similarly to a bulk PhTC that amplifies free-space harmonics inside the momentum bandgap. But note that the amplification only occurs at the surface and not
in free space, and therefore the metasurface is not acting entirely like a ``bulk'' PhTC. 
Interestingly, in this case, the modulation frequency is only 2/3 of the incident frequency, which is three times reduced compared to usual bulk PhTCs that are modulated at twice the incident frequency \cite{lustig2018topological}.
Such configuration has also been observed in bulk PhTCs \cite{zurita2009reflection, park2021spatiotemporal}. 


\subsection*{Metasurface design  at microwave frequencies}

As a proof of concept, we design a metasurface for operation at microwave frequencies and experimentally explore the momentum bandgap for surface waves. Figure~\ref{fig: bigfigure}A shows the unit cell of the suggested metasurface. The capacitive impenetrable metasurface is implemented by a periodic arrangement of metallic strips in the $xy$-plane over a grounded substrate~\cite{sievenpiper1999high}. 
On the top side, two metal strips are separated by a gap that provides capacitive coupling for TE-polarized surface waves.
To modulate the effective capacitance of the unit cell in time, the metallic strips are connected through vias and a varactor diode located below the bottom metallization layer.   
The cathode of the diode is biased with a static voltage $V_{\rm dc}=3.5$~V to provide the average capacitance value of the varactor $C_{\rm av}=4$~pF (different from the effective capacitance of the unit cell).
Using full-wave simulations in CST Studio Suite, the effective \textit{surface} capacitance of the time-invariant unit cell is estimated as  $C_0=0.95$~pF. At frequencies below 1~GHz, the time-invariant unit cell supports only one eigenmode in TE polarization (see the band structure in Fig.~S2). 
The surface wave is most efficiently transmitted at around $f_{\rm s}=870$~MHz, with the wavelength of $\lambda_s\approx 4 D_x$, as shown in Fig.~S6.


The metasurface of finite size comprises eight described unit cells along the $x$-direction.
Since the metasurface operates in TE polarization (electric field along the $y$-direction), the periodic arrangement of the unit cells along the $y$-direction is conveniently emulated by placing vertical conducting walls (parallel to the $xz$-plane) at $y=\pm D_y/2$ (see the geometry of the metasurface in Fig.~S3).
The sinusoidal modulation signal $V_{\rm mod}(t)=V_0\sin(2\pi f_{\rm m}t)$ with $f_{\rm m}=2f_{\rm s}$ was injected into a $1\times 4$ power divider to modulate the capacitance of the four unit cells. 
Numerical simulations show that modulating only four out of eight unit cells (approximately one-wavelength area) is sufficient to achieve amplification for surface waves (Fig.~S5). The amplification factor is largely insensitive to the phase of the incident wave (see related discussions on phase sensitivity in the supplementary section~S7). 



\subsection*{Experimental demonstration}

\begin{figure*}
\includegraphics[width=0.99\linewidth]{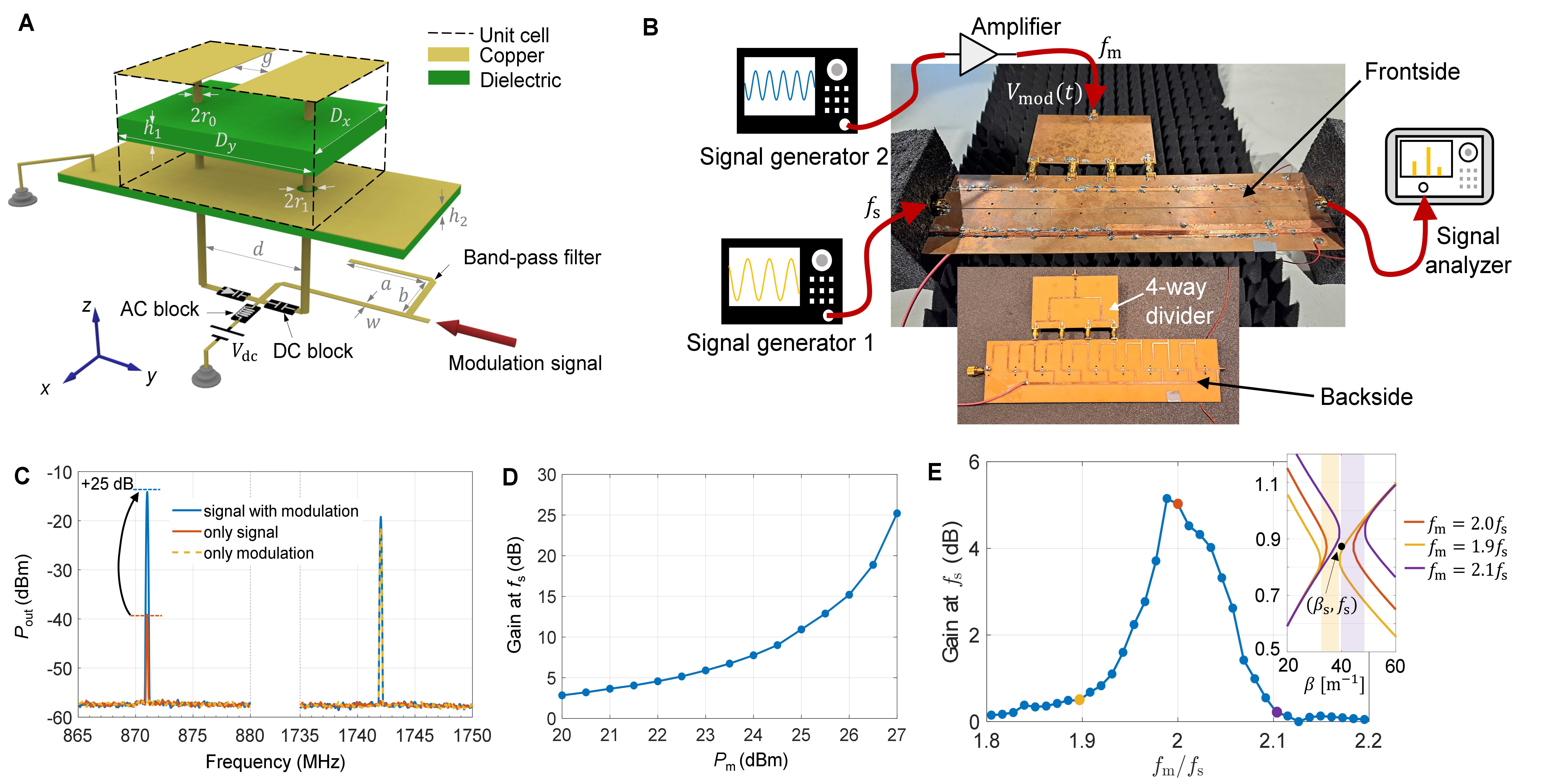}
\caption{\textbf{Experimental results for surface-wave excitation.} (A) Unit cell of the metasurface-based PhTC (denoted in a dashed frame). The unit cell is periodically repeated in the $xy$-plane. Here, for clearer illustration, the three layers with metallic patterns (upper layer with strips, middle ground plane, and lower layer with circuitry) are visually separated from the dielectric substrates.
(B) Experiment setup. The black modules at two terminals are microwave absorbers. Since the output voltage of the signal generator~2 is limited, a radio-frequency voltage amplifier with 40~dBm gain is connected to improve the modulation amplitude. (C) Frequency spectrum of the received power at the output port for three different  scenarios. (D) Dependence of the amplification gain at $f_{\rm s}=871$~MHz on the power of the  modulation signal. (E) Amplification gain of the surface wave at $f_{\rm s}=871$~MHz versus the normalized modulation frequency $f_{\rm m}/f_{\rm s}$. The modulation strength is kept constant at 22.5~dBm. The black point represents the incident surface harmonics ($\beta_{\rm s}, f_{\rm s}$). The band structure is obtained numerically, for the extracted value of the effective capacitance of the modulated metasurface   $C(t)=C_0 [1-0.25\cos(2\pi f_{\rm m}t)]$, where $C_0=0.95$~pF.}
\label{fig: bigfigure}
\end{figure*}

The fabricated metasurface-based PhTC is shown in Fig.~\ref{fig: bigfigure}B.
Before switching on the modulation, the operating frequency of the metasurface is determined as $f_{\rm 0}=871$~MHz by measuring the maximum transmission for surface waves traveling through the metasurface (Fig.~S6). 

Next, we investigate the surface wave transmission when temporal modulations are turned on. The experimental setup can be seen in Fig.~\ref{fig: bigfigure}B.
The output port is connected to a spectrum analyzer. The modulation is generated by an independent generator 2 with a frequency twice the signal frequency $f_{\rm m}=2f_{\rm s}=1742~{\rm MHz}$ to ensure that the incident surface wave falls into the stopband. The modulation signal enters into a 4-way microstrip power divider which is connected to the four unit cells of the metasurface. 

Figure~\ref{fig: bigfigure}C shows the spectrum of the signal received at Port~2 inside the momentum bandgap. In the absence of modulation, there is only the signal frequency harmonic with power $P_{\rm out}(f_{\rm s})=-39.14$~dBm. When the modulation signal is exerted at the middle port with $+27$~dBm  power, the output power of the fundamental frequency components increases to $P_{\rm out}(f_{\rm s})=-14.13$~dBm, that is, with $25$~dB gain. 
The amplification of the surface wave occurs because the incident signal wave has a wavenumber inside the momentum bandgap induced by the temporal modulation. 
When no signal wave is present and the modulation is on, the output port detects only frequency harmonic $f_{\rm m}$ (see yellow dashed line in Fig.~\ref{fig: bigfigure}C), which means that the metasurface does not reach  the unstable regime of parametric  self-oscillations. 
While working in the parametric-amplification regime, the PhTC exhibits a finite  gain: Radiation losses due to a finite-size geometry compensate for the exponential signal growth. The power of the surface wave at $f_{\rm s}$ received at Port~2 increases exponentially with the strength of the temporal modulations $P_{\rm m}=V_0^2/2 R_{\rm s}$ (here $R_{\rm s}=50$~Ohm is the source resistance), as shown in Fig.~\ref{fig: bigfigure}D. Further increase of the modulation strength would lead to parametric self-oscillation whose amplitude would be limited due to the departing from the linear regime of the varactor diode~\cite{Gonorovsky}.

It is also important to experimentally confirm the non-amplification effect for excitations outside the momentum bandgap. To depart from the bandgap, it is sufficient to shift the modulation frequency away from the value $f_{\rm m}=2f_{\rm s}$ (keeping constant $f_{\rm s}$). 
In this way, the band structure of the PhTC  will be modified (the band structure is uniquely determined by the metasurface structure and the modulation frequency), and the signal surface harmonic with parameters $(\beta_{\rm s}, f_{\rm s})$ will no longer belong to the momentum bandgap. The inset in Fig.~\ref{fig: bigfigure}E depicts    numerically calculated bandgaps for three scenarios with different modulation frequencies. The black point denotes the signal surface harmonic. As is seen, for the modulation frequency $f_{\rm m}=1.9f_{\rm s}$~MHz and $f_{\rm m}=2.1f_{\rm s}$~MHz, the incident wave resides right at the edge of the bandgap. The primary plot in Fig.~\ref{fig: bigfigure}E indicates the measured amplification gain of the PhTC at the fixed signal frequency $f_{\rm s}=871$~MHz versus changing modulation frequency $f_{\rm m}$. 
When the modulation frequency is close to twice the incident frequency, amplification is maximum because the wave is in the center of the bandgap. As the modulation frequency reduces (increases), the bandgap is shifted to the left (right) in the momentum axis, and the amplification rate decreases. 
The asymmetry in the  amplification rate with respect to   the central frequency in Fig.~\ref{fig: bigfigure}E is caused by the asymmetric dispersion of the metasurface (as can be seen from Fig.~S2).


It is worth mentioning that the amplification effect of incident surface wave obtained in both simulations and experiments is nearly phase-insensitive.  Therefore, the PhTC does not require phase synchronization between incident and modulation signals. As seen in Fig.~\ref{fig: bigfigure}B, there is no synchronization between signal and modulation sources.  Such a property fundamentally differs from conventional degenerate parametric amplifiers operating in the stable regime (section S8). The phase-insensitivity of observed phenomena in the experimental metasurface-based PhTC is additionally demonstrated in Fig.~S11. %

\begin{figure*}
\includegraphics[width=0.99\linewidth]{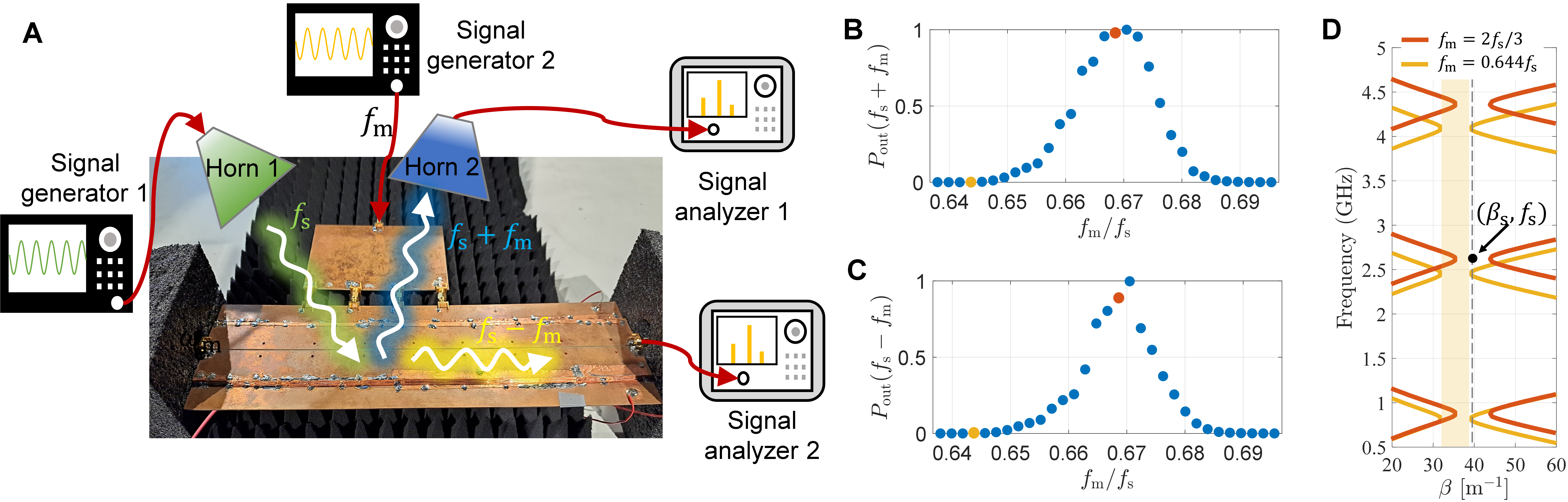}
\caption{\textbf{Experimental results for free-space-wave excitation. }(A) Experimental setup for free-space excitation of the momentum bandgap. 
 The transmitting antenna (Horn 1) is oriented at $\theta_{\rm i} \approx 45^\circ$. The modulation power generated by Signal Generator~2 is 22.5~dBm. Incident signal frequency $f_{\rm s}$ is fixed. 
(B) Measured power of induced  harmonic at $f_{\rm s}+f_{\rm m}$ versus the modulation frequency. The power value is normalized with respect to the peak power.   The modulation frequency is normalized by the  signal frequency $f_{\rm s}$.
(C) Same for harmonic $f_{\rm s}-f_{\rm m}$.  (D) Band structure for the cases of  $f_{\rm m}=2f_{\rm s}/3$ (red curve) and $f_{\rm m}=0.644f_{\rm s}$ (yellow curve). The black point represents the incident surface harmonics ($\beta_{\rm s}, f_{\rm s}$). The band structure is obtained numerically, for the extracted value of the effective capacitance of the modulated metasurface   $C(t)=C_0 [1-0.25\cos(2\pi f_{\rm m}t)]$, where $C_0=0.95$~pF. }
\label{fig: freespace}
\end{figure*}

Next, we experimentally show that the momentum bandgap can be directly accessed by free-space excitation. Figure~\ref{fig: freespace}A shows the modified experimental setup. The incident beam is generated by a   directive horn antenna (Horn 1) fed by a Signal Generator~1 operating at $f_{\rm s}=2613$~MHz (note that the signal frequency in this experiment is different from that in the experiment presented in Fig.~\ref{fig: bigfigure}). 
The scattered field contains infinite numbers of harmonics $f_s$, $f_s \pm f_{\rm m}$, $f_s \pm 2f_{\rm m}$, etc. We only measure the lowest-order scattered harmonic by Signal analyzers~1 and~2, $f_{\rm s}+ f_{\rm m}$ and $f_{\rm s}- f_{\rm m}$,  corresponding to free space and surface modes, respectively.   
We did not measure the specularly reflected harmonic at $f_{\rm s}$  due to the presence  of   strong parasitic coupling (crosstalk) between transmitting and receiving antennas.
In Fig. 3B and 3C, it is shown that when the modulation frequency is close to $2f_{\rm s}/3$, the detected signals at $f_{\rm s}+ f_{\rm m}$ and $f_{\rm s}- f_{\rm m}$ are nearly maximal (see red points in the plots). The reason for this effect is the fact that  the excitation harmonic is located in the center of the bandgap (see the red curves in Fig.~3D). 
To verify that the amplification is due to the excitation being in the momentum bandgap, the modulation frequency was swept around $f_{\rm m}={2f_{\rm s}}/{3}$. The sweep leads to the shift of the bandgap.  
As the modulation frequency reduces to $f_{\rm m}=0.644f_{\rm s}$, the signal wave (black point in Fig. 3D) falls out of the bandgap (the shaded yellow region in Fig. 3D), and the detected signals by  both analyzers are minimized, as shown by the yellow point of Figs.~3B and 3C. 

\section*{Discussion}



This work demonstrates that a temporally modulated metasurface offers equivalent wave effects to bulk three-dimensional PhTCs. 
We have experimentally verified the existence of momentum bandgap in such two-dimensional PhTCs and strong wave amplification inside it. Furthermore, it is shown that a metasurface-based PhTC can be excited externally by plane waves, amplifying both surface-bounded and free-space harmonics.
The proposed metasurface platform can notably simplify the construction and exploration of PhTCs, providing great convenience for exploring photonic time and space-time  crystals~\cite{caloz2019spacetime,  sharabi2022spatiotemporal}. Although  in this work for simplicity we chose the unit-cell size along the $x$-direction (see Fig.~\ref{fig: bigfigure}) of the order of the quarter-wavelength of the surface modes, this parameter can be notably increased for creating arbitrary space-time modulations or decreased for modeling more accurately pure space-uniform (temporal-only) modulations.
The space-time modulations can be synthesized with our metasurface by adding phase shifters to the power divider.

From the application point of view, the demonstrated metasurface PhTCs can become an important contribution to future communications, serving as an amplifier for surface-wave signals that are known to  suffer from severe losses.  In particular, recently suggested surface-wave-assisted reconfigurable intelligent surfaces~\cite{shojaeifard2022mimo}  would enormously benefit from the amplification provided by the introduced metasurfaces. 
Such surfaces are predicted to operate as smart radio environments, playing a central role in future   wireless communications. The typical operational range of such systems is 24~GHz or lower, therefore,  similar varactor-based metasurfaces can be exploited.
Furthermore, the proposed concept is general (based on the surface-impedance description) and not limited to the microwave frequency regime. Potentially, it can be   scaled up  
to sub-THz frequencies using tunable 2D-materials and optical frequencies using nonlinear effects~\cite{chen2020all}.


\section*{Materials and Methods}
\subsection*{Numerical simulation of time-varying capacitive boundary}
The field simulations in Figs.~\ref{fig: bbigfigure}D, 1E, 1F, and 1G are obtained from COMSOL Multiphysics. In COMSOL, the transient solver (temw) is chosen to simulate the time-varying boundary. The boundary is modelled by defining a surface current density which is written in the form of Eq.~S3.  The surface wave is excited by defining the oscillating electric field (along $y$-direction) at the left edge of the simulation domain in Figs.~1D. The plane wave excitation in Figs.~1F and 1G are defined by a time-varying electric field on the top boundary of the simulation domain.
\subsection*{Metasurface design, simulation, and fabrication}
In each meta-atom, to prevent the crosstalk of the surface wave with the modulation wave, an L-shaped microstrip band-pass filter is connected to the bias circuit. 
The structural parameters of the meta-atom shown in Fig.~\ref{fig: bigfigure}A are $g=1.0$~mm, $r_1=2r_0=2.0$~mm,  $D_x=40$~mm, $D_y=50$~mm, $h_1=2.4$~mm, $h_2=1.0$~mm, $d=9.0$~mm, $a=25.6$~mm, $b=21.0$~mm, and $w=2.0$~mm. The dielectric substrate is FR4 with permittivity $\epsilon_r=4.2(1-j0.025)$. 
The actual metasurface consists of 8 meta-atoms. The numerical simulation of the actual metasurface is carried out in CST Studio Suite 2021 using EM \& Circuit Co-simulation method. The electromagnetic
(EM) structure is modeled in the EM interface, and with circuit components connected in schematic interface. More details on the full-wave simulation method is available in section~S5.
The metasurface is fabricated using printed circuit board (PCB) technology.
\subsection*{Measurement}
In the measurement setup shown in Fig.~2B and 3A, signal generator 1 is Keysight MXG Analog Signal Generator N5181B, signal generator 2 is Rohde \& Schwarz SMIQ-06B, and the signal analyzer is Tektronix RSA 5126B. The RF amplifier is Milmega AS0822-8L. During the measurement, wave absorbers are positioned at the left, right, and bottom sides of the metasurface to prevent strong wave reflection.  The DC bias lines are soldered at the bottom side of the metasurface. The measurement is carried out in the anechoic chamber.  All the data are captured randomly.

\section*{Supplementary Materials}
This PDF file includes \\
Figs.~S1 to S12\\
Sections~S1 to S9
\\\\
Other Supplementary Material for this
manuscript includes the following:\\
Movies S1 and S2

\bibliography{references}

\bibliographystyle{Science}

\section*{Acknowledgments}
The authors would like to thank Dr.~Nam Van Ha and Dr. Linping Feng for helpful discussions on the experiment, and Mr.~Quangang Chen and Ms.~Yining Liu for their kind support in the experiment.
\section*{Funding}
This work has been supported by the Academy of Finland, project 330260. X. W. and C. R. receive support from the Helmholtz program “Materials Systems Engineering” (MSE).~S. F. receives MURI grants from the U. S. Air Force Office of Scientific Research (Grant No. FA9550-18-1-0379 and FA9550-21-1-0244).
\section*{Author Contributions}
M.S.M. conceived the idea of analyzing eigenmodes and dispersion curves of time-varying boundaries. V.S.A. and X.W. developed the idea further to the realization of two-dimensional photonic time crystals based on such time-varying boundaries. X.W. and M.S.M. did the initial theoretical verification. X.W. performed the other theoretical calculations and all the numerical simulations. X.W. and V.S.A. made the experimental measurements. S.A.T., S.F., and C.R. supervised the work. All the authors contributed to the discussions of the results and the manuscript preparation.
\section*{Competing interests}
 The authors declare that they have no competing interests. 
\section*{Data and materials availability}
 All data needed to evaluate the conclusions in the paper are present in the paper and/or the Supplementary Materials.




\clearpage

\end{document}